# Proposal of room-temperature diamond maser

Liang Jin[1], Matthias Pfänder[2], Nabeel Aslam[2], Sen Yang[2], Jörg Wrachtrup[2] & Ren-Bao Liu[1,*]

1. Department of Physics & Centre for Quantum Coherence, The Chinese University of Hong Kong, Shatin, New Territories, Hong Kong, China

2. 3rd Institute of Physics, University of Stuttgart, 70569 Stuttgart, Germany

* Corresponding author. rbliu@phy.cuhk.edu.hk

**Abstract:**

**Lasers have revolutionized optical science and technology, but their microwave counterpart, maser, has not realized its great potential due to its demanding work conditions (high-vacuum for gas maser and liquid-helium temperature for solid-state maser). Room-temperature solid-state maser is highly desirable, but under such conditions the lifetimes of emitters (usually electron spins) are usually too short (~ns) for population inversion. The only room-temperature solid-state maser is based on a pentacene-doped *p*-terphenyl crystal, which has long spin lifetime (~0.1 ms). This maser, however, operates only in the pulse mode and the material is unstable. Here we propose room-temperature maser based on nitrogen-vacancy (NV) centres in diamond, which feature long spin lifetimes at room temperature (~10 ms), high optical pump efficiency, and material stability. We demonstrate that under readily accessible conditions, room-temperature diamond maser is feasible. Room-temperature diamond maser may facilitate a broad range of microwave technologies.**

Maser[1], the prototype of laser in the microwave waveband, has important applications[2-5] such as in ultrasensitive magnetic resonance spectroscopy, astronomy observation, space communication, radar, and high-precision clocks. Such applications, however, are hindered by the demanding operation conditions (high vacuum for gas maser[6] and liquid-helium temperature for solid-state maser[7]). Room-temperature solid-state maser is highly desirable, but the lifetime of emitters (electron spins) under such conditions is usually too short (~ns[8]) for population inversion under practical pump conditions. Long spin lifetime in organic materials has enabled a room-temperature solid-state maser[9], which, however, operates only at the pulse mode with a low repetition rate (~1 Hz). Stimulated microwave emission from silicon-vacancy defects in silicon carbide has been demonstrated at room temperature[10] but maser has not been achieved. Here we propose room-temperature maser based on nitrogen-vacancy (NV) centre electron spins in diamond[11]. NV centres feature the longest known spin lifetime at room temperature among all solid-state systems (~10 ms[12,13], versus ~0.1 ms in organic materials[9] and ~0.5 ms in silicon carbide[14]) and high optical pumping efficiency (~$10^6$ s$^{-1}$ [15], versus ~$10^3$-$10^5$ s$^{-1}$ in organic materials[9] and ~$10^5$ s$^{-1}$ in silicon carbide[16]). Our numerical simulation demonstrates that under readily accessible conditions (cavity $Q$ factor ~$10^5$, pump rate ~$10^5$ s$^{-1}$), room-temperature diamond maser with coherence time ~60,000 s (or linewidth ~5 μHz) is feasible. For a room-temperature microwave amplifier, the noise temperature is as low as ~0.1 K (versus ~1 K for the state-of-the-art ruby amplifier working at liquid-helium temperature[3]). Room-temperature diamond maser provides a new **stage for studying macroscopic quantum coherence in spin ensembles and may facilitate a broad range of microwave technologies.**

Nitrogen-vacancy centre spins in diamond[11] have been extensively studied for quantum information processing[11] and sensing[17-19], due to their long coherence time at room temperature and high efficiencies of initialization by optical pumping and readout via photon detection. In particular, coupling between ensemble NV centre electron spins and microwave cavities have been demonstrated for quantum information storage and retrieval in hybrid quantum systems[20]. Enhanced quantum coherence of NV centre ensemble has been observed in the strong coupling regime[21]. On the basis of these well-established works, here we propose a new class of applications of NV centre in diamond, namely, room-temperature solid-state maser and microwave amplifier.

The key to maser – laser in the microwave waveband (frequency in 0.3~300 GHz or wavelength in 1 m~1 mm) – is population inversion of the emitters and macroscopic coherence among the microwave photons. Population inversion requires a spin relaxation rate lower than the pump rate. This sets the bottleneck in room-temperature solid-state maser, as the spin relaxation times in solids are usually extremely short (~ns[8]) at room temperature due to rapid phonon scattering[7]. The spin relaxation induced by phonon scattering can be largely suppressed in light-element materials (such as organic materials) where the spin-orbit coupling is weak. Actually, the only room-temperature solid-state maser is based on a pentacene-doped *p*-terphenyl molecular crystal where the spin lifetime can reach 135 μs at room temperature[9]. However, the active spins in pentacene-doped *p*-terphenyl are intermediate metastable states instead of the ground states. Such an energy level structure greatly reduces the optical pumping efficiency and requires high pump laser power (>200 W). An even more serious issue is the material instability of *p*-terphenyl molecular crystals. The organic maser operates only at the pulse mode with a rather low repetition rate (~1 Hz) [9]. Another candidate under consideration is silicon vacancy ($V_{Si}$) centre in silicon carbide[10]. Very recently, it has been clarified that $V_{Si}$ centres have a spin-3/2 ground state, which allows population inversion by optical pump[10,14]. The spin lifetime is quite long (~0.5 ms) at room temperature[14]. Stimulated microwave emission from $V_{Si}$ centres has been observed. Maser from the silicon carbide spins, however, is still elusive. The challenges include the complexity of defects in the compound material and the difficulty of engineering the $V_{Si}$ centres.

NV centres in diamond, as established by researches in the context of quantum computing, possess all features necessary for room-temperature solid-state maser. NV centre spins have the longest known lifetime (~10 ms) at room temperature[12,13] among all solid-state spins (~100 times longer than in pentacene-doped *p*-terphenyl[9] and ~20 times longer than in silicon carbide[10,14]). The spin is a triplet (spin-1) in the ground state and can be rapidly pumped by optical laser to the $m_s = 0$ ground state, with a pump rate as high as ~$10^6$ s$^{-1}$ [15] (versus ~$10^3$-$10^5$ s$^{-1}$ in organic materials[9] and ~$10^5$ s$^{-1}$ in silicon carbide[16]). Therefore the population inversion can be easily achieved if a magnetic field is applied to shift the $m_s = 0$ ground state to above another spin state. Furthermore, large single crystals of diamond can be grown by chemical vapour deposition (CVD)[22] and the NV centres can be created by ion implantation with tuneable density[23]. Strong coupling between large ensembles of NV centre spins and high-quality microwave cavities has been demonstrated[24,25]. The good thermal conductivity and material stability of diamond are also advantageous for maser. All these features suggest NV centres in diamond a superb candidate for room-temperature solid-state maser.

We consider an ensemble of NV centre spins in diamond resonantly coupled to a high quality Febry-Pérot microwave cavity (Fig. 1a, see Supplementary Information Note A for details). Note that many other types of microwave cavities[9,10,24-27] may be employed for implementing the proposal in this paper. The spin sublevels $|m_s\rangle$ ($m_s = 0$ or $m_s = \pm 1$) of the NV triplet ground state have a zero-field splitting about 2.87 GHz between $|0\rangle$ and $|\pm 1\rangle$[11] (Fig. 1b). The NV centres can be optically pumped to the state $|0\rangle$[11]. A moderate external magnetic field (>1000 Gauss) splits the $|\pm 1\rangle$ and shifts the $|-1\rangle$ state to below $|0\rangle$ so that the spins can be inverted by optical pump. The transition frequency $\omega_S$ between the spin ground state $|g\rangle \equiv |-1\rangle$ and the spin exited state $|e\rangle \equiv |0\rangle$ is tuned resonant with the microwave cavity frequency $\omega_c$.

The maser is driven by coupling between the cavity mode and the spins. The Hamiltonian of the coupled spin-cavity system is $H_I = \sum_{j=1}^{N} g_j \left( \hat{a} \hat{s}_j^+ + \hat{a}^\dagger \hat{s}_j^- \right)$, where $\hat{a}$ annihilates a microwave cavity photon, $\hat{s}_j^+ \equiv |e\rangle_{jj}\langle g|$ is the raising operator of the $j$-th spin, $\hat{s}_j^- = \left( \hat{s}_j^+ \right)^\dagger$, and $g_j$ is the coupling constant. Without changing the essential results, we assume the spin-photon coupling is a constant, i.e., $g_j = g$ and write the Hamiltonian as $H_I = g \left( \hat{a} \hat{S}_+ + \hat{a}^\dagger \hat{S}_- \right)$ with the collective operators $\hat{S}_\pm \equiv \sum_{j=1}^{N} \hat{s}_j^\pm$, which satisfy the commutation relation $\left[ \hat{S}_+, \hat{S}_- \right] = \sum_{j=1}^{N} \left( |e\rangle_{jj}\langle e| - |g\rangle_{jj}\langle g| \right) \equiv \hat{S}_z$. When masing

occurs, the spin polarization (or population inversion) $S_z \equiv \langle \hat{S}_z \rangle$ is a macroscopic number [$\sim O(N)$] while the fluctuation $\delta \hat{S}_z \equiv \hat{S}_z - S_z \sim O(N^{1/2})$ is much smaller. Therefore, $\hat{b}^\dagger \equiv \hat{S}_-/\sqrt{S_z}$ can be interpreted as the creation operator of a collective magnon mode with $[\hat{b}, \hat{b}^\dagger] \cong 1$. In the masing state, both the photon and the magnon modes, coherently coupled to each other, have macroscopic amplitudes. With the number of coherent magnons $n_S \equiv \langle \hat{b}^\dagger \hat{b} \rangle = \langle \hat{S}_+ \hat{S}_- \rangle / S_z \sim O(N)$, the spins are in a macroscopic quantum superposition state maintained by the masing process.

Now we describe the masing process. A prerequisite of maser is to invert the spin population (see Fig. 1b). The optical pumping rate $w$ can be tuned by varying the pump laser intensity, up to $\sim 10^6$ s$^{-1}$ [15]. The cavity mode has a decay rate determined by the cavity $Q$ factor, $\kappa_c = \omega_c/Q$, due to photon leakage and coupling to input/output channels. The decay of the magnon mode is caused by various mechanisms. First, the spin relaxation ($T_1$ process caused by phonon scattering and resonant interaction between spins) contributes a decay rate $\gamma_{eg} = 1/T_1$. Second, the individual spins experience local field fluctuations due to interaction with nuclear spins, coupling to other NV and nitrogen (P1) centre electron spins, and fluctuation of the zero-field splitting[20]. Such local field fluctuations induce random phases to individual spins, making the bright magnon mode decay to other modes at a rate $2/T_2^*$, where $T_2^*$ is the dephasing time of the spin ensembles. Finally the optical pump of NV centres, being incoherent, also induces decay of the magnon mode. The magnon decay rate induced by the incoherent pump is $qw$ ($q \approx 16$, see Supplementary Information Note A for estimation), which is larger than the spin pump rate $w$ due to multiple excitation and relaxation pathways (Fig. 1c). The total decay rate of the magnon mode is thus $\kappa_S = qw + 2/T_2^* + \gamma_{eg}$. The quantum dynamics of the coupled magnons and photons is described by the quantum Langevin equations[28] for the photon and magnon operators $\hat{a}$ and $\hat{S}_\pm$, the spin populations $\hat{N}_{e/g} \equiv \sum_{j=1}^{N} |e/g\rangle_{jj} \langle e/g|$, and $\hat{S}_z$ (see Methods for details).

For a specific system, we consider a diamond of volume $V_{NV} = 3\times3\times0.5$ mm$^3$ with natural abundance (1.1%) of $^{13}$C nuclear spins, P1 centre concentration about 5 ppm, and the NV centre concentration $\rho_{NV} = 10^{17}$ cm$^{-3}$ (for N-to-NV conversion efficiency ~10%). The ensemble spin decoherence time is $T_2^* = 0.5$ μs [20] (see Supplementary Information Note A for estimation). Considering the four orientations of NV centres and three nuclear spin states of $^{14}$N, the number of NV centres coupled to the cavity mode is estimated to be $N = \rho_{NV} V_{NV}/12 \approx 0.375\times10^{14}$. The external magnetic field 2100 Gauss results in $\omega_S/2\pi \approx 3$ GHz. The microwave cavity has length $L \approx 50$ mm and has its frequency resonant with the magnon, i.e., $\omega_c = \omega_S$. The spin-photon coupling is about $g/2\pi = 0.02$ Hz for the effective cavity mode volume $V_{eff} \approx 2$ cm$^3$ [24]. At room temperature ($T$=300 K), the phonon scattering dominates the spin relaxation and $\gamma_{eg} \approx 200$ s$^{-1}$ [13]. The number of thermal photons inside the cavity is $n_{th} \approx 2,100$.

The quantum Langevin equations can be solved at steady-state masing. When maser occurs, the quantum operators can be approximated as their expectation values, i.e., $\hat{S}_\pm \approx S_\pm$, $\hat{a} \approx a$, $\hat{N}_{e/g} \approx N_{e/g}$, and $\hat{S}_z \approx S_z$. By dropping the small quantum fluctuations, we reduce the quantum Langevin equations to classical equations for the expectation values (see Methods for details). Under the resonant condition ($\omega_S = \omega_c$), the steady-state solution is

$$S_z = \kappa_S \kappa_c / (4g^2),$$
$$S_- = i\sqrt{S_z \left( \frac{w - \gamma_{eg}}{2\kappa_S} N - \frac{w + \gamma_{eg}}{2\kappa_S} S_z \right)}, \qquad (1)$$
$$a = \sqrt{\frac{w - \gamma_{eg}}{2\kappa_c} N - \frac{w + \gamma_{eg}}{2\kappa_c} S_z}.$$

Note that to have a population inversion (spin polarization) $S_z \sim O(N)$, the pump rate should scale with the total number of spins as $w \sim O(N)$. The fact that the photon number $|a|^2 = \frac{w - \gamma_{eg}}{2\kappa_c} N - \frac{w + \gamma_{eg}}{2\kappa_c} S_z > 0$ leads to the masing condition

$$\kappa_c < \frac{4g^2}{\kappa_S} \frac{w - \gamma_{eg}}{w + \gamma_{eg}} N. \qquad (2)$$

First, the pump rate needs to be greater than the spin relaxation rate for population inversion ($w > \gamma_{eg}$). Second, the cavity $Q$ factor has to be above a threshold to have a sufficient number of photons for sustaining the macroscopic quantum coherence. Stronger spin-photon coupling ($g$), longer magnon lifetime ($\kappa_S^{-1}$), or larger number ($N$) of spins can reduce this threshold of cavity $Q$ factor. Finally, the magnon decay rate $\kappa_S$ should be kept below the maximal collective emission rate of photons $4Ng^2/\kappa_c$, otherwise over pumping would fully polarize the spins, making the spin-spin correlation vanish ($S_z \to N$ and $S_- \to 0$).

Emergence of macroscopic quantum coherence is evidenced by macroscopic values of the spin polarization, the photon number and the magnon amplitude under the masing condition. We calculated these values by using the higher order equations of the correlation functions (see Methods and Supplementary Information Note B for details), which apply to both maser and incoherent emission. The spin polarization $S_z$, the microwave output $P_{\text{out}} = \hbar\omega_c \cdot \kappa_c \langle \hat{a}^\dagger \hat{a} \rangle$, and the spin coherence $\langle \hat{S}_+ \hat{S}_- \rangle$ (shown in Fig. 2a-c) are consistent with results obtained from equation (1) when the pump rate and the cavity $Q$ factor are above the masing threshold (white curve in the figures). It is clearly seen that the output power increases dramatically when the pump rate is above the spin relaxation rate (population inverted) and the cavity $Q$ factor is above the masing threshold (Fig. 2b). Since the pump rate $w \sim O(N)$, the output power scales with the number of spins by $P_{\text{out}} \approx \hbar\omega_c \cdot w(N - S_z)/2 \sim O(N^2)$, which demonstrates the superradiant nature of the maser. The fact that $\langle \hat{S}_+ \hat{S}_- \rangle \gg N_e$ unambiguously evidences phase correlation among the large spin ensemble established by cavity photons in the masing region (Fig. 2c). The optimal pump condition for spin correlation is determined by maximizing

$$\langle \hat{S}_+ \hat{S}_- \rangle = S_z \left( \frac{w - \gamma_{eg}}{2\kappa_S} N - \frac{w + \gamma_{eg}}{2\kappa_S} S_z \right). \tag{3}$$

With the assumption $\gamma_{eg} \ll w, 1/T_2^*$, the spin-spin correlation reaches its maximum value $\langle \hat{S}_+ \hat{S}_- \rangle \approx N^2/(8q)$ at the optimal pump rate $w_{\text{opt}}^{\text{max-corr}} \approx 2Ng^2/(q\kappa_c)$, where the spin polarization $S_z \approx N/2$ and the maser power $P_{\text{out}} \approx \hbar\omega_c \cdot N^2 g^2/(2q\kappa_c)$ (See Supplementary Information Note D for derivation).

The maser linewidth is determined by the correlation of the phase fluctuations of photons or equivalently that of magnons. The coherence time (see Supplementary Information Note C for derivation) is obtained as

$$T_{\text{coh}} = 4\left(\kappa_c^{-1} + \kappa_S^{-1}\right)(n_c + n_S)/n_{\text{incoh}}, \qquad (4)$$

where $n_c = \langle \hat{a}^\dagger \hat{a} \rangle$ is the photon number, $n_S = \langle \hat{S}_+ \hat{S}_- \rangle / S_z$ is the magnon number, and $n_{\text{incoh}} = n_{\text{th}} + N_e/S_z$ includes the thermal photon number ($n_{\text{th}}$) and the incoherent magnon number ($\langle \hat{S}_+ \hat{S}_- \rangle / S_z = \sum_{j=1}^{N} \langle \hat{s}_j^+ \hat{s}_j^- \rangle / S_z = N_e/S_z$ if the correlation between different spins is neglected). The coherence time is greatly enhanced under the masing condition (Fig. 2d). For NV centres, the magnon decay rate $\kappa_S > 10^6$ s$^{-1}$ while for a good microwave cavity ($Q > 10^5$) the photon decay rate $\kappa_c < 3 \times 10^4$ s$^{-1}$. Thus the photon number $n_c = n_S \kappa_S / \kappa_c$ is much greater than the magnon number and the macroscopic quantum coherence is mainly maintained by the photons in the cavity. For a cavity with quality factor $Q = 10^5$ and a pump rate $w = 10^5$ s$^{-1}$ (marked by a black cross in Fig. 2), which are readily accessible in experiments, the coherence time is as long as $6.0 \times 10^4$ s. The optimal pump condition for long coherence time can be obtained from equation (4). In the good-cavity or large ensemble limit where $2Ng^2/\kappa_c \gg 1/T_2^*$ and at room temperature where $n_{\text{incoh}} \approx n_{\text{th}}$, the optimal pump rate for maximum coherence time is close to that for maximum spin correlation, i.e., $w_{\text{opt}}^{\text{max-corr}} \approx 2Ng^2/(q\kappa_c)$, and the optimal coherence time reaches

$$T_{\text{coh}}^{\text{opt}} \approx 2N^2 g^2 / \left(q n_{\text{th}} \kappa_c^3\right). \qquad (5)$$

The maximum coherence time scales with the spin number and the cavity $Q$ factor as $T_{\text{coh}}^{\text{opt}} \propto N^2 Q^3$ (see Supplementary Information Note D). Note that the quantum coherence sustained by active masing (with pump) has much longer lifetime than spin coherence protected by passive coupling to the cavity (without pump)[21].

The long coherence time of maser is a useful resource for quantum technologies, such as ultrasensitive active magnetometry[29]. When the external magnetic field or the mirror position (cavity size) is changed such that the spin transition frequency is shifted away from the exact resonance with the cavity photon, the masing frequency will be dragged to $\omega = (\kappa_c \omega_S + \kappa_S \omega_c)/(\kappa_c + \kappa_S)$ (see Supplementary Information Note B)[30]. The ultralong coherence time of the maser means ultranarrow linewidth and hence ultrasensitivity to the external magnetic field and the mirror position (see Fig. 2e,f). The sensitivity of a magnetic field with frequency $\leq (\kappa_c + \kappa_S)/2$ is estimated to be $\delta B \sqrt{t_m} = \gamma_{\text{NV}}^{-1}(1 + \kappa_S/\kappa_c)\sqrt{2 T_{\text{coh}}^{-1}}$ for measurement time $t_m$ (see Supplementary Information Note C & E for derivation), where $\gamma_{\text{NV}}/2\pi = 2.8\,\text{MHz} \cdot \text{Gauss}^{-1}$ is the NV centre gyromagnetic ratio. The magnetic field sensitivity can reach up to $1\,\text{pT} \cdot \text{Hz}^{-1/2}$ for $Q = 10^5$ and $w = 10^5\,\text{s}^{-1}$ at room temperature. Under the same condition, the sensitivity to the cavity mirror position, $\delta x \sqrt{t_m} = (L/\omega_c)(1 + \kappa_c/\kappa_S)\sqrt{2 T_{\text{coh}}^{-1}}$, is about $16\,\text{fm} \cdot \text{Hz}^{-1/2}$. For higher cavity $Q$ factor, the position sensitivity is greatly enhanced while the magnetometry sensitivity is reduced due to the frequency dragging effect (see Fig. 2e,f). The sensitivities to the magnetic field and the mirror positions set the requirements on stability of the setup for maintaining the long coherence time of the maser.

The coupled spin-cavity system can be configured as a microwave amplifier when the spin population is inverted but the cavity lifetime is below the masing threshold. To study the amplification, we calculate the spin inversion, the output microwave power, and the noise temperature with a weak microwave input (see Methods and Supplementary Information Note F for details). As shown in Fig. 3, when the spins are inverted ($S_z > 0$) but the cavity $Q$ factor is below the masing threshold [inequality (2) violated], the system linearly amplifies the microwave signal. For readily accessible parameters ($w=10^5$ s$^{-1}$ & $Q=4\times10^4$), the gain is about 25 dB and the noise temperature is as low as 152 mK, which indicates single-photon noise level.

## METHODS SUMMARY

The theoretical study is based on the standard Langevin equations[28]

$$\frac{d\hat{N}_e}{dt} = +w\hat{N}_g - \gamma_{eg}\hat{N}_e + ig\left(\hat{a}^\dagger \hat{S}_- - \hat{S}_+ \hat{a}\right) + \hat{F}_e,$$

$$\frac{d\hat{N}_g}{dt} = -w\hat{N}_g + \gamma_{eg}\hat{N}_e - ig\left(\hat{a}^\dagger \hat{S}_- - \hat{S}_+ \hat{a}\right) + \hat{F}_g,$$

$$\frac{d\hat{S}_-}{dt} = -i\omega_S \hat{S}_- - \frac{\kappa_S}{2}\hat{S}_- + ig\left(\hat{N}_e - \hat{N}_g\right)\hat{a} + \hat{F}_S,$$

$$\frac{d\hat{a}}{dt} = -i\omega_c \hat{a} - \frac{\kappa_c}{2}\hat{a} - ig\hat{S}_- + \hat{F}_c,$$

(6)

where $\hat{F}_{c/S/e/g}$ is the noise operator that causes the decay of the photons (c), the magnons (S), the population in the excited state (e), or that in the ground state (g). Note that the total spin number is written as an operator $\hat{N}$ to take into account the fluctuation due to population of the third spin state $|+1\rangle$. The population fluctuation, however, has no effect on the phase fluctuation of the maser (see Supplementary Information Note C).

By replacing the operators with their expectation values, we obtain the mean-field equations for the maser at the steady-state

$$0 = wN_g - \gamma_{eg}N_e + ig(a^*S_- - S_+a),$$
$$0 = i(\omega - \omega_S)S_- - \frac{\kappa_S}{2}S_- + igS_z a, \quad (7)$$
$$0 = i(\omega - \omega_c)a - \frac{\kappa_c}{2}a - igS_-,$$

from which the masing frequency, the field amplitudes, and the spin polarization can be straightforwardly calculated.

The coherence time and linewidth are calculated using the spectrum of the phase fluctuations. The Langevin equations are linearized for the fluctuations, which are much smaller than the expectation values at steady-state. The linearized equations are

$$\frac{d\delta \hat{N}_e}{dt} = +w\delta \hat{N}_g - \gamma_{eg}\delta \hat{N}_e + ig\left(S_-\delta \hat{a}^\dagger - S_+\delta \hat{a}\right) + ig\left(a^*\delta \hat{S}_- - a\delta \hat{S}_+\right) + \hat{F}_e,$$
$$\frac{d\delta \hat{N}_g}{dt} = -w\delta \hat{N}_g + \gamma_{eg}\delta \hat{N}_e - ig\left(S_-\delta \hat{a}^\dagger - S_+\delta \hat{a}\right) - ig\left(a^*\delta \hat{S}_- - a\delta \hat{S}_+\right) + \hat{F}_g,$$
$$\frac{d\delta \hat{S}_-}{dt} = -\frac{\kappa_S}{2}\delta \hat{S}_- + igS_z\delta \hat{a} + iga\left(\delta \hat{N}_e - \delta \hat{N}_g\right) + \hat{F}_S, \quad (8)$$
$$\frac{d\delta \hat{a}}{dt} = -\frac{\kappa_c}{2}\delta \hat{a} - ig\delta \hat{S}_- + \hat{F}_c.$$

By Fourier transform of these equations, the spectrum of the phase noise can be calculated and hence the maser linewidth is determined.

To investigate the correlations in both the masing and the incoherent emission regimes, we derive the equations for the correlation functions and take the expectation values of the relevant operators. That leads to

$$\frac{d\langle \hat{N}_e \rangle}{dt} = +w\langle \hat{N}_g \rangle - \gamma_{eg}\langle \hat{N}_e \rangle + ig\left(\langle \hat{a}^\dagger \hat{S}_- \rangle - \langle \hat{S}_+ \hat{a} \rangle\right),$$

$$\frac{d\langle \hat{N}_g \rangle}{dt} = -w\langle \hat{N}_g \rangle + \gamma_{eg}\langle \hat{N}_e \rangle - ig\left(\langle \hat{a}^\dagger \hat{S}_- \rangle - \langle \hat{S}_+ \hat{a} \rangle\right),$$

$$\frac{d\langle \hat{a}^\dagger \hat{S}_- \rangle}{dt} = -\frac{\kappa_S + \kappa_c}{2}\langle \hat{a}^\dagger \hat{S}_- \rangle + ig\left[\left(1 - \frac{1}{N}\right)\langle \hat{S}_+ \hat{S}_- \rangle + \langle \hat{N}_e \rangle + \langle \hat{a}^\dagger \hat{a} \rangle\langle \hat{S}_z \rangle\right], \quad (9)$$

$$\frac{d\langle \hat{S}_+ \hat{S}_- \rangle}{dt} = -\kappa_S \langle \hat{S}_+ \hat{S}_- \rangle - ig\langle \hat{S}_z \rangle\left(\langle \hat{a}^\dagger \hat{S}_- \rangle - \langle \hat{S}_+ \hat{a} \rangle\right),$$

$$\frac{d\langle \hat{a}^\dagger \hat{a} \rangle}{dt} = -\kappa_c \langle \hat{a}^\dagger \hat{a} \rangle - ig\left(\langle \hat{a}^\dagger \hat{S}_- \rangle - \langle \hat{S}_+ \hat{a} \rangle\right) + \kappa_c n_{th}.$$

Here to make the equations close, we have used the approximation

$\langle \hat{a}^\dagger \hat{a} \hat{S}_z \rangle \approx \langle \hat{a}^\dagger \hat{a} \rangle \langle \hat{S}_z \rangle$, $\langle \hat{a}^\dagger \hat{S}_z \hat{S}_- \rangle \approx \langle \hat{S}_z \rangle \langle \hat{a}^\dagger \hat{S}_- \rangle$, and $\langle \hat{S}_+ \hat{S}_z \hat{a} \rangle \approx \langle \hat{S}_z \rangle \langle \hat{S}_+ \hat{a} \rangle$, neglecting

the higher-order correlations, which is well justified for Gaussian fluctuations.

To investigate the microwave amplifier, we solve the mean-field equations with a steady-state input $s_{in} e^{-i\omega_{in} t}$ as

$$0 = wN_g - \gamma_{eg} N_e + ig(a^* S_- - S_+ a),$$

$$0 = i(\omega_{in} - \omega_S)S_- - \frac{\kappa_S}{2} S_- + igS_z a,$$

$$0 = i(\omega_{in} - \omega_c)a - \frac{\kappa_c}{2} a - igS_- + \sqrt{\kappa_{ex}} s_{in}, \quad (10)$$

$$s_{out} = s_{in} - \sqrt{\kappa_{ex}} a,$$

from which power gain $G = |s_{out}|^2 / |s_{in}|^2$ is obtain. Maximum gain $G \gg 1$ is possible under the resonant condition $\omega_{in} = \omega_c = \omega_S$, but will be reduced to $O(1)$ at off-resonance, i.e., $|\omega_{in} - \omega_{S,c}|/\kappa_{S,c} \gtrsim 1$. The intrinsic noise temperature[7] of the diamond maser is given by

$$T_n = (1 - G^{-1})\left[\frac{L_{dB}}{G_{dB}} T + \left(1 + \frac{L_{dB}}{G_{dB}}\right)\frac{N_e}{S_z}\frac{\hbar \omega_c}{k_B}\right], \quad (11)$$

where $T$ is the environment temperature, $L_{dB} = -10\log_{10} e^{-\kappa_c \tau_{rt}}$ is the cavity power loss in decibel during the time of a photon roundtrip $\tau_{rt} = 2L/c$, and $G_{dB} = 10\log_{10} G$.

**Full methods, supplementary figures and associated references are available in the online Supplementary Information.**

**Acknowledgements** This work was supported by Hong Kong Research Grants Council Project No. 14303014 & The Chinese University of Hong Kong Focused Investments Scheme.

**Author Contributions** R.B.L. conceived the idea and supervised the project, R.B.L. & L.J. formulated the theory, R.B.L., L.J., M.P., N.A., S.Y. & J.W. designed the physical system, L.J. carried out the study, R.B.L. & L.J. wrote the paper, and all authors commented on the manuscript.

**Author Information** Correspondence and requests for materials should be addressed to R.B.L. (rbliu@phy.cuhk.edu.hk).

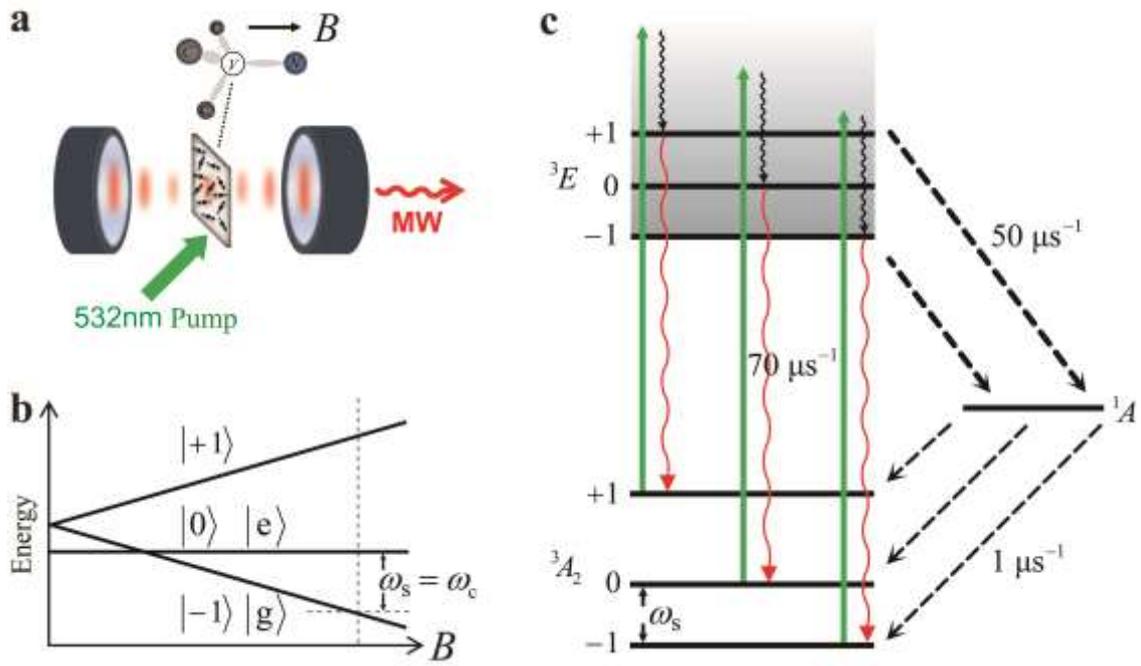

**Figure 1 | Scheme of room-temperature diamond maser. a**, The system for maser. A diamond sample is fixed inside a high-quality Febry-Pérot microwave cavity. A magnetic field is applied along the NV axis, which is set parallel to the cavity axis. The NV centres are pumped by a 532 nm laser (green arrow). **b**, The energy levels of an NV centre spin as functions of a magnetic field *B*. The zero-field splitting at *B*=0 is about 2.87 GHz. The magnetic field is set such that the transition frequency $\omega_s$ between the states $|-1\rangle$ ($|g\rangle$) and $|0\rangle$ ($|e\rangle$) is resonant with the cavity mode $\omega_c$. **c**, The pump scheme. After the spin-conserving optical excitation by a 532 nm laser (green arrows), the excited state $^3E$ can directly return to the ground state $^3A_2$ via spin-conserving photon emission at a rate ~70 μs$^{-1}$, but the excited states $|m_s = \pm 1\rangle$ can also decay to the metastable singlet state $^1A$ via inter-system crossing at a rate ~50 μs$^{-1}$ and the singlet state decays to the three different ground states at a rate ~1 μs$^{-1}$ in each pathway.

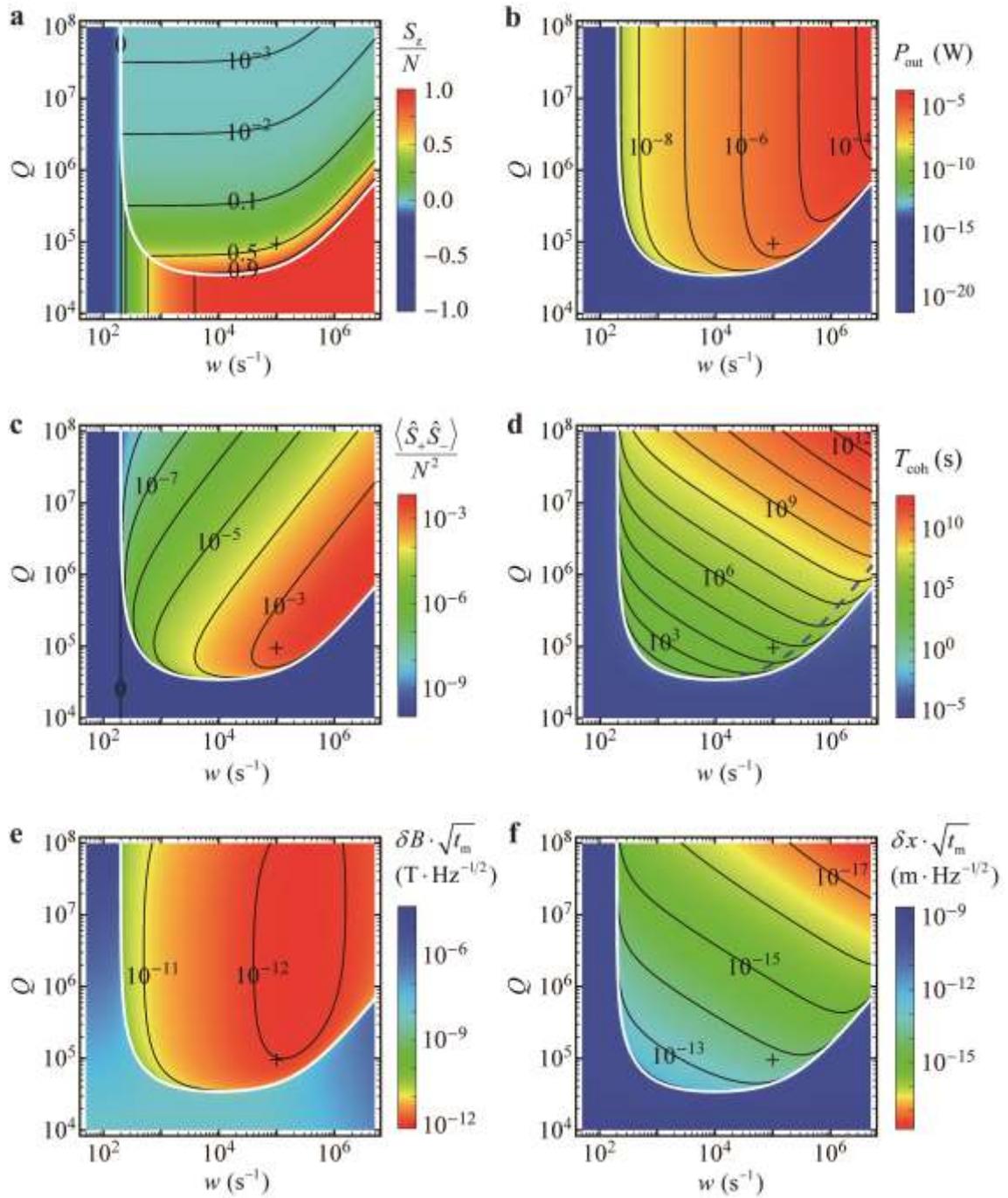

**Figure 2 | Room-temperature masing of NV centre spins in diamond.** Contour plots of **a**, the spin polarization $S_z$, **b**, the output power $P_{out}$, **c**, the collective spin-spin correlation $\langle \hat{S}_+ \hat{S}_- \rangle$, **d**, the macroscopic quantum coherence time $T_{coh}$, **e**, the sensitivity on external magnetic field, and **f**, the sensitivity on cavity size, as functions of the cavity

*Q* factor and the pump rate *w*. The masing threshold for *w* and *Q* is indicated in the figures by the white curves. The blue dashed curve in **d** shows the optimal pump condition for maximum coherence time. The crosses in the figures mark the point for $Q = 10^5$ and $w = 10^5$ s$^{-1}$. The parameters are such that $\omega_c/2\pi = \omega_s/2\pi = 3$ GHz, $g/2\pi = 0.02$ Hz, $T_2^* = 0.5$ μs, $N = 0.375 \times 10^{14}$, *T*=300 K, and $\gamma_{eg} = 200$ s$^{-1}$.

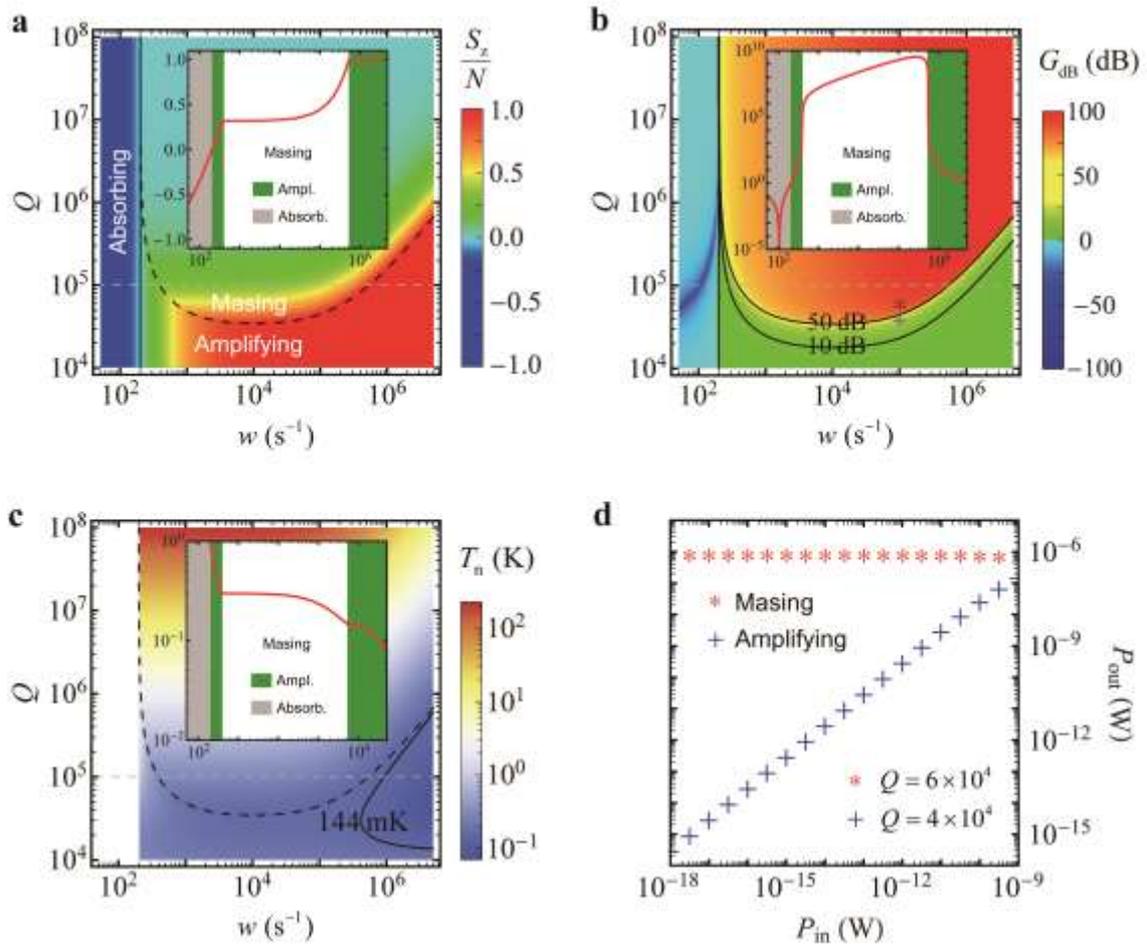

**Figure 3 | Room-temperature diamond microwave amplifier.** Contour plots of **a**, the spin inversion $S_z$, **b**, the power gain $G$, and **c**, the noise temperature $T_n$, as functions of the pump rate and the cavity $Q$ factor for input microwave power $P_{in}$=1 fW. For $w < \gamma_{eg} = 200$ s$^{-1}$, the system is in the absorbing region (population not inverted). For $w > \gamma_{eg} = 200$ s$^{-1}$ and the cavity $Q$ factor below/above masing threshold (the dashed curve), the system operates in the amplifying/masing mode. The insets show dependence on the pump rate for a fixed $Q=10^5$ (the absorbing, amplifying and masing regions marked as the grey, green and white). **d**, Output power as a function of the input microwave power, in the amplifying and masing regions (corresponding to the parameters marked by cross and star in **b**, respectively). The amplification is linear in a

large range of input power in the amplifying region (see Supplementary Fig. S3 for more information). The parameters are the same as in Fig. 2.

# SUPPLEMENTARY INFORMATION for

# Proposal of room-temperature diamond maser

**Note A. Details of the system**

**NV centre concentration**—We consider a chemical vapour deposition (CVD) bulk diamond sample with size of $V_{NV} = 3 \times 3 \times 0.5 \text{ mm}^3$. The concentration of P1 centre (single substitutional nitrogen centres) is about 5 ppm, and the NV centres concentration is about $\rho_{NV} = 10^{17} \text{ cm}^{-3}$ for the $N \rightarrow NV^-$ conversion efficiency ~10% (conversion efficiency up to about 30% is available with the state-of-the-art technology [S1]).

**Spin-photon coupling**—We consider a microwave cavity with resonant frequency $\omega_c / 2\pi = 3 \text{ GHz}$. The cavity length is $L \approx 50 \text{ mm}$, and the cavity mirrors' radii of curvature are about 25.1 mm, which result in a Gaussian microwave beam waist ≈7 mm for the resonant cavity mode and an effective cavity mode volume $V_{eff} \approx 2 \text{ cm}^3$. Assuming that the NV centre axis is parallel to the cavity axis, we estimate the spin-photon coupling to be $\frac{g}{2\pi} = \frac{\gamma_{NV}}{2\pi} \sqrt{\frac{\mu_0 \hbar \omega_c}{2 V_{eff}}} \approx 0.02 \text{ Hz}$ [S2], where $\mu_0$ is the vacuum permittivity.

**NV centre number**—The number of NV centre spins $N$ coupled to the cavity photons is estimated as follows. Considering four orientations of the NV centres and the hyperfine interaction with the nitrogen nuclear spin (which splits the transition into three resonances with only one resonant with the cavity mode), the number of NV centres effectively interacting with the cavity mode is estimated to be $N = \rho_{NV} V_{NV} / 12 = 0.375 \times 10^{14}$. The small population of the $m_s = +1$ spin state leads to the noise of the total spin number $\hat{N}$. At steady-state, we take $N$ as the average value of $\hat{N}$.

**Spin dephasing time**—The dephasing of an ensemble of NV centres is mainly caused by dipolar interaction with the P1 centre electron spins, the hyperfine interaction with $^{13}$C nuclear spins, and the zero-field splitting (ZFS) fluctuations. Interactions with the natural abundance (1.1%) of $^{13}$C nuclear spins contributes a dephasing rate ~$10^6$ s$^{-1}$ [S3, S4], the dipolar interaction with 5 ppm concentration of P1 centre electron spins contributes ~$10^6$ s$^{-1}$ [S5-S8], and the zero-field splitting fluctuation induces a dephasing rate ~$10^6$ s$^{-1}$ [S9, S10]. We estimate the dephasing time $T_2^* = 0.5\,\mu$s for the ensemble of NV centres (for comparison, $T_2^* \approx 0.43\,\mu$s was measured for the ensemble of NV centres in a diamond with natural abundance of $^{13}$C and P1 centre concentration ~20 ppm [S9], and $T_2^* \approx 0.20\,\mu$s for P1 centre concentration ~40 ppm [S10]).

**Spin-lattice relaxation time**—The spin-lattice relaxation dominates the NV centre spin relaxation at room temperature. We take $\gamma_{eg} = 200$ s$^{-1}$ ($T_1 = 5$ ms) at $T$=300 K [S11].

**Optical pump process**—The NV centre electronic spins are optically pumped into the $m_s$=0 ground state (Fig. 1c in the main text). A 532 nm green laser excites the NV centre triplet ground state $^3A_2$ to an excited vibronic state, which decays to the triplet excited state $^3E$. The $m_s$=0 excited state almost fully decays to the ground state via spin-conserving photon emission. However, the $m_s$=±1 excited state can decay to the ground state either by spin-conserving photon emission or by spin non-conserving relaxation via the metastable singlet state $^1A$ due to the inter-system crossing (ISC) [S12-S14]. The spin conserving emission and the ISC have similar rates. The metastable singlet state relaxes to the three ground states at roughly equal rate of 1 μs$^{-1}$ [S14]. Under continuous optical pumping, the $m_s$=−1 ground state ($|g\rangle$) is pumped to the $m_s$=0 ground state ($|e\rangle$) with a success probability of about 1/4 in each excitation event. Considering effective pump into $|e\rangle$ state at rate $w$, the excitation rate is about 4$w$. The photon emission processes to the $m_s$=−1 and $m_s$=0 ground states have rates 2$w$ and 4$w$, respectively. All these pump and decay processes destroy the NV centre spin ensemble coherence. Therefore the decay of the magnons due to the optical pump has a rate about 16$w$.

The total magnon decay rate, including the contributions of spin relaxation, spin dephasing, and optical pump, is thus obtained as $\kappa_S = qw + 2/T_2^* + \gamma_{eg}$, where $q\approx16$.

**Note B. Langevin equations and the steady-state solution**

The quantum Langevin equations read [S15]

$$\frac{d\hat{N}_e}{dt} = +w\hat{N}_g - \gamma_{eg}\hat{N}_e + ig\left(\hat{a}^\dagger \hat{S}_- - \hat{S}_+ \hat{a}\right) + \hat{F}_e, \tag{S1a}$$

$$\frac{d\hat{N}_g}{dt} = -w\hat{N}_g + \gamma_{eg}\hat{N}_e - ig\left(\hat{a}^\dagger \hat{S}_- - \hat{S}_+ \hat{a}\right) + \hat{F}_g, \tag{S1b}$$

$$\frac{d\hat{S}_-}{dt} = -i\omega_S \hat{S}_- - \frac{\kappa_S}{2}\hat{S}_- + i\hat{g}(t)\hat{S}_z \hat{a} + \hat{F}_S, \tag{S1c}$$

$$\frac{d\hat{a}}{dt} = -i\omega_c \hat{a} - \frac{\kappa_c}{2}\hat{a} - ig\hat{S}_- + \hat{F}_c, \tag{S1d}$$

where $\hat{N}_e - \hat{N}_g = \hat{S}_z$, $\hat{N}_e + \hat{N}_g = \hat{N}$, the noise operator $\hat{F}_{e/g}$ is the population fluctuation in the spin state $|e/g\rangle$, $\hat{F}_e + \hat{F}_g$ is the fluctuation of the NV centre electron spin number due to population in the $m_s=+1$ state and other intermediate states, and $\hat{F}_{S/c}$ is the magnon/photon noises.

The mean-field theory is well justified when masing occurs, since the fluctuations are much smaller than the expectation values of the operators in equation (S1). We write the operators as the sum of their expectation values and small fluctuations, i.e., $\hat{N}_{e/g} = N_{e/g} + \delta\hat{N}_{e/g}$, $\hat{S}_\pm = S_\pm e^{\pm i\omega t} + \delta\hat{S}_\pm$, $\hat{a} = ae^{-i\omega t} + \delta\hat{a}$, where $\omega$ is the steady-state frequency of the maser. The steady-state mean-field equations are

$$0 = wN_g - \gamma_{eg}N_e + ig(a^*S_- - S_+ a), \tag{S2a}$$

$$0 = i(\omega - \omega_S)S_- - \frac{\kappa_S}{2}S_- + igS_z a, \tag{S2b}$$

$$0 = i(\omega - \omega_c)a - \frac{\kappa_c}{2}a - igS_-, \tag{S2c}$$

From equations (S2b) and (S2c), we obtain

$$S_- = \frac{-igS_z a}{i(\omega - \omega_S) - \frac{\kappa_S}{2}}, \quad a = \frac{igS_-}{i(\omega - \omega_c) - \frac{\kappa_c}{2}}, \tag{S3}$$

which give the spin polarization (population inversion)

$$S_z = \frac{[\kappa_c - 2i(\omega - \omega_c)][\kappa_S - 2i(\omega - \omega_S)]}{4g^2}. \tag{S4}$$

The inversion $S_z$ being real requires $(\omega-\omega_c)\kappa_S+(\omega-\omega_S)\kappa_c=0$, which determines the maser frequency $\omega=\dfrac{\kappa_c\omega_S+\kappa_S\omega_c}{\kappa_c+\kappa_S}$. Inserting $\omega$ into equation (S4), we get

$$S_z=\frac{\kappa_S\kappa_c}{4g^2}\left(1+\delta_{cs}^2\right), \tag{S5}$$

where $\delta_{cs}=2(\omega_c-\omega_S)/(\kappa_c+\kappa_S)$ is the frequency mismatch between the spin and cavity [S16]. At resonance $\omega_S=\omega_c=\omega$, the solutions are $\kappa_c\kappa_S=4g^2S_z$, $\kappa_c a=-2igS_-$, and the magnon and photon numbers satisfy $\kappa_c n_c=\kappa_S\left(S_+S_-/S_z\right)=\kappa_S n_S$.

To calculate the spin correlations, we use

$$\frac{d\langle\hat{N}_e\rangle}{dt}=w\langle\hat{N}_g\rangle-\gamma_{eg}\langle\hat{N}_e\rangle+ig\left(\langle\hat{a}^\dagger\hat{S}_-\rangle-\langle\hat{S}_+\hat{a}\rangle\right), \tag{S6a}$$

$$\frac{d\langle\hat{N}_g\rangle}{dt}=-w\langle\hat{N}_g\rangle+\gamma_{eg}\langle\hat{N}_e\rangle-ig\left(\langle\hat{a}^\dagger\hat{S}_-\rangle-\langle\hat{S}_+\hat{a}\rangle\right), \tag{S6b}$$

$$\frac{d\langle\hat{a}^\dagger\hat{S}_-\rangle}{dt}=-\frac{\kappa_S+\kappa_c}{2}\langle\hat{a}^\dagger\hat{S}_-\rangle+ig\left[(1-\frac{1}{N})\langle\hat{S}_+\hat{S}_-\rangle+\langle\hat{N}_e\rangle+\langle\hat{a}^\dagger\hat{a}\rangle\langle\hat{S}_z\rangle\right], \tag{S6c}$$

$$\frac{d\langle\hat{S}_+\hat{S}_-\rangle}{dt}=-\kappa_S\langle\hat{S}_+\hat{S}_-\rangle-ig\langle\hat{S}_z\rangle\left(\langle\hat{a}^\dagger\hat{S}_-\rangle-\langle\hat{S}_+\hat{a}\rangle\right), \tag{S6d}$$

$$\frac{d\langle\hat{a}^\dagger\hat{a}\rangle}{dt}=-\kappa_c\langle\hat{a}^\dagger\hat{a}\rangle-ig\left(\langle\hat{a}^\dagger\hat{S}_-\rangle-\langle\hat{S}_+\hat{a}\rangle\right)+\kappa_c n_{th}, \tag{S6e}$$

where symbol $\langle\cdots\rangle$ denotes the expectation values of the operators. We have used the approximations $\langle\hat{a}^\dagger\hat{a}\hat{S}_z\rangle\approx\langle\hat{a}^\dagger\hat{a}\rangle\langle\hat{S}_z\rangle$, $\langle\hat{a}^\dagger\hat{S}_z\hat{S}_-\rangle\approx\langle\hat{S}_z\rangle\langle\hat{a}^\dagger\hat{S}_-\rangle$, and $\langle\hat{S}_+\hat{S}_z\hat{a}\rangle\approx\langle\hat{S}_z\rangle\langle\hat{S}_+\hat{a}\rangle$, dropping the higher order correlations [S17, S18]. The steady-state expectation values is calculated from equation (S6) by setting $d\langle\hat{N}_{e/g}\rangle/dt=d\langle\hat{a}^\dagger\hat{S}_-\rangle/dt=d\langle\hat{S}_+\hat{S}_-\rangle/dt=d\langle\hat{a}^\dagger\hat{a}\rangle/dt=0$, which leads to

$$0=wN_g-\gamma_{eg}N_e+ig\left(\langle\hat{a}^\dagger\hat{S}_-\rangle-\langle\hat{S}_+\hat{a}\rangle\right), \tag{S7a}$$

$$0=-\frac{\kappa_S+\kappa_c}{2}\langle\hat{a}^\dagger\hat{S}_-\rangle+ig\left[(1-\frac{1}{N})\langle\hat{S}_+\hat{S}_-\rangle+N_e+\langle\hat{a}^\dagger\hat{a}\rangle S_z\right], \tag{S7b}$$

$$0=-\kappa_S\langle\hat{S}_+\hat{S}_-\rangle-igS_z\left(\langle\hat{a}^\dagger\hat{S}_-\rangle-\langle\hat{S}_+\hat{a}\rangle\right), \tag{S7c}$$

$$0=-\kappa_c\langle\hat{a}^\dagger\hat{a}\rangle-ig\left(\langle\hat{a}^\dagger\hat{S}_-\rangle-\langle\hat{S}_+\hat{a}\rangle\right)+\kappa_c n_{th}. \tag{S7d}$$

The spin-spin correlations $\langle \hat{S}_+ \hat{S}_- \rangle$, the populations of the two spin states $N_{e/g}$, the population inversion (spin polarization) $S_z$, and the photon number $n_c$ can be obtained from equation (S7). The incoherent thermal photons are far less than the stimulated emission photons and mainly affect the linewidth of the microwave field. Above the threshold ($w > \gamma_{eg}$), we can safely neglect $n_{th}$ in equation (S7). Solving equation (S7) yields a quadratic equation about $S_z$, that is,

$$S_z^2 - \left[\frac{w-\gamma_{eg}}{w+\gamma_{eg}}N + \frac{\frac{\kappa_S+\kappa_c}{4g^2} + \frac{1}{w+\gamma_{eg}}}{(1-N^{-1})\frac{1}{\kappa_S}+\frac{1}{\kappa_c}}\right]S_z + \frac{w-\gamma_{eg}}{w+\gamma_{eg}}N\frac{\frac{\kappa_S+\kappa_c}{4g^2} - \frac{1}{w-\gamma_{eg}}}{(1-N^{-1})\frac{1}{\kappa_S}+\frac{1}{\kappa_c}} = 0. \quad (S8)$$

Since $N \gg 1$ and $\frac{\kappa_S+\kappa_c}{4g^2} \gg \frac{1}{w \pm \gamma_{eg}}$, we have $(1-N^{-1}) \approx 1$ and $\frac{\kappa_S+\kappa_c}{4g^2} \pm \frac{1}{w \pm \gamma_{eg}} \approx \frac{\kappa_S+\kappa_c}{4g^2}$. Equation (S8) is further reduced to

$$S_z^2 - \left(\frac{w-\gamma_{eg}}{w+\gamma_{eg}}N + \frac{\kappa_S \kappa_c}{4g^2}\right)S_z + \frac{w-\gamma_{eg}}{w+\gamma_{eg}} \times \frac{N\kappa_S \kappa_c}{4g^2} = 0. \quad (S9)$$

The population inversion in the masing region is given by the stable solution

$$S_z = \frac{\kappa_S \kappa_c}{4g^2}. \quad (S10)$$

And in the amplifying region, the stable solution is $S_z = \frac{w-\gamma_{eg}}{w+\gamma_{eg}}N$ (details of amplifying is discussed in Supplementary Information Note F).

**Note C. Quantum diffusion and coherence time**

The linearized quantum Langevin equations for the fluctuations are (second order terms such as $\delta\hat{a}^{\dagger}\delta\hat{S}_{-}$ dropped)

$$\frac{d\delta\hat{N}_{e}}{dt} = +w\delta\hat{N}_{g} - \gamma_{eg}\delta\hat{N}_{e} + ig(S_{-}\delta\hat{a}^{\dagger} - S_{+}\delta\hat{a}) + ig(a^{*}\delta\hat{S}_{-} - a\delta\hat{S}_{+}) + \hat{F}_{e}, \quad \text{(S11a)}$$

$$\frac{d\delta\hat{N}_{g}}{dt} = -w\delta\hat{N}_{g} + \gamma_{eg}\delta\hat{N}_{e} - ig(S_{-}\delta\hat{a}^{\dagger} - S_{+}\delta\hat{a}) - ig(a^{*}\delta\hat{S}_{-} - a\delta\hat{S}_{+}) + \hat{F}_{g}, \quad \text{(S11b)}$$

$$\frac{d\delta\hat{S}_{-}}{dt} = -\frac{\kappa_{S}}{2}\delta\hat{S}_{-} + igS_{z}\delta\hat{a} + iga(\delta\hat{N}_{e} - \delta\hat{N}_{g}) + \hat{F}_{S}, \quad \text{(S11c)}$$

$$\frac{d\delta\hat{a}}{dt} = -\frac{\kappa_{c}}{2}\delta\hat{a} - ig\delta\hat{S}_{-} + \hat{F}_{c}. \quad \text{(S11d)}$$

Here we have assumed the resonant condition $\omega_{c} = \omega_{S}$. By Fourier transform $\delta\hat{a}(\Omega) = (2\pi)^{-1/2}\int_{-\infty}^{+\infty} e^{i\Omega t}\delta\hat{a}(t)$ for $\delta\hat{a}(t)$ and other fluctuation operators, we obtain the noise operator in the frequency domain. Without loss of generality, we set the intra-cavity field $a$ as a real number. Under the resonant condition the magnon amplitude $S_{-}$ is purely imaginary.

The quantum Langevin equations in frequency domain yield

$$-i\Omega\delta\hat{S}_{-}(\Omega) = -\frac{\kappa_{S}}{2}\delta\hat{S}_{-}(\Omega) + iga[\delta\hat{N}_{e}(\Omega) - \delta\hat{N}_{g}(\Omega)] + igS_{z}\delta\hat{a}(\Omega) + \hat{F}_{S}(\Omega), \quad \text{(S12a)}$$

$$-i\Omega\delta\hat{S}_{+}(\Omega) = -\frac{\kappa_{S}}{2}\delta\hat{S}_{+}(\Omega) - iga^{*}[\delta\hat{N}_{e}(\Omega) - \delta\hat{N}_{g}(\Omega)] - igS_{z}\delta\hat{a}^{\dagger}(\Omega) + \hat{F}_{S}^{\dagger}(\Omega), \quad \text{(S12b)}$$

$$-i\Omega\delta\hat{a}(\Omega) = -\frac{\kappa_{c}}{2}\delta\hat{a}(\Omega) - ig\delta\hat{S}_{-}(\Omega) + \hat{F}_{c}(\Omega), \quad \text{(S12c)}$$

$$-i\Omega\delta\hat{a}^{\dagger}(\Omega) = -\frac{\kappa_{c}}{2}\delta\hat{a}^{\dagger}(\Omega) + ig\delta\hat{S}_{+}(\Omega) + \hat{F}_{c}^{\dagger}(\Omega). \quad \text{(S12d)}$$

The phase fluctuation of magnons is

$$\delta\hat{\phi}_{S} = [\delta\hat{S}_{-}(\Omega) + \delta\hat{S}_{+}(\Omega)] = \frac{i(\frac{\kappa_{c}}{2} - i\Omega)[\hat{F}_{S}(\Omega) + \hat{F}_{S}^{\dagger}(\Omega)] - \frac{\kappa_{c}\kappa_{S}}{4g}[\hat{F}_{c}(\Omega) - \hat{F}_{c}^{\dagger}(\Omega)]}{\Omega(\frac{\kappa_{c} + \kappa_{S}}{2} - i\Omega)}, \quad \text{(S13)}$$

and that of photons is

$$\delta\hat{\phi}_c(\Omega) = -i[\delta\hat{a}(\Omega) - \delta\hat{a}^\dagger(\Omega)] = \frac{(\frac{\kappa_S}{2} - i\Omega)[\hat{F}_c(\Omega) - \hat{F}_c^\dagger(\Omega)] - ig[\hat{F}_S(\Omega) + \hat{F}_S^\dagger(\Omega)]}{\Omega(\frac{\kappa_c + \kappa_S}{2} - i\Omega)}. \quad (S14)$$

With the correlation functions of the field noise operators, $\langle \hat{F}_c(\Omega)\hat{F}_c(\Omega')\rangle = 0$, $\langle \hat{F}_c^\dagger(\Omega)\hat{F}_c(\Omega')\rangle = \kappa_c n_{th}\delta(\Omega+\Omega')$, and $\langle \hat{F}_c(\Omega)\hat{F}_c^\dagger(\Omega')\rangle = \kappa_c(1+n_{th})\delta(\Omega+\Omega')$, we obtain $[\hat{F}_c(\Omega) - \hat{F}_c^\dagger(\Omega)][\hat{F}_c(\Omega') - \hat{F}_c^\dagger(\Omega')] = -\kappa_c(1+2n_{th})\delta(\Omega+\Omega')$. Similarly, the noise operators for the spins satisfy $[\hat{F}_S(\Omega) + \hat{F}_S^\dagger(\Omega)][\hat{F}_S(\Omega') + \hat{F}_S^\dagger(\Omega')] = N\kappa_S \delta(\Omega+\Omega')$.

The phase noise spectrum $S_c(\Omega) = \langle \delta\hat{\phi}_c(\Omega)\delta\hat{\phi}_c(-\Omega)\rangle$ is calculated as

$$\frac{S_c(\Omega)}{4n_c} = \frac{\left(\frac{\kappa_c+\kappa_S}{2}\right)^2}{\Omega^2\left[\left(\frac{\kappa_c+\kappa_S}{2}\right)^2 + \Omega^2\right]} \left[\frac{g^2 N\kappa_S + \left(\frac{\kappa_S^2}{4} + \Omega^2\right)\kappa_c(1+2n_{th})}{4n_c\left(\frac{\kappa_c+\kappa_S}{2}\right)^2}\right]. \quad (S15)$$

Since it is much less than the cavity and magnon decay rates, the maser linewidth is determined by the phase noises at low frequencies, $\Omega \ll (\kappa_c + \kappa_S)/2$. So the frequency fluctuation of the maser can be approximated as $\left\langle \frac{d}{dt}\delta\hat{\phi}_c(t)\frac{d}{dt'}\delta\hat{\phi}_c(t')\right\rangle = \frac{1}{2\pi}\int_{-\infty}^{+\infty} e^{-i\Omega(t-t')}\Omega^2 S_c(\Omega)d\Omega \approx \gamma_{ST}\delta(t-t')$, where

$$\gamma_{ST} = \left(\frac{N_e}{S_z} + n_{th}\right)\frac{\kappa_c}{2n_c}\left(\frac{\kappa_S}{\kappa_c+\kappa_S}\right)^2 = n_{incoh}\cdot\frac{1}{n_c+n_S}\cdot\frac{1}{(\kappa_S/2)^{-1}+(\kappa_c/2)^{-1}}, \quad (S16)$$

is the Schawlow-Townes diffusion coefficient, with the incoherent magnons and photons number $n_{incoh} = N_e/S_z + n_{th}$. For $t \gg 2/(\kappa_c+\kappa_S)$, the phase correlation is $\left\langle \delta\hat{\phi}_c^2(t)\right\rangle = \int_0^t dt'\int_0^t dt''\left\langle \delta\dot{\hat{\phi}}_c(t')\delta\dot{\hat{\phi}}_c(t'')\right\rangle = \gamma_{ST}\cdot t$. Neglecting the amplitude fluctuations (which is negligible at steady-state) and assuming a Gaussian statistics for the phase fluctuations, the spectrum of the cavity photon field is

$$\langle \hat{a}(\Omega)\hat{a}(-\Omega)\rangle = \int_{-\infty}^{+\infty} dt\, e^{i\Omega t}\langle \hat{a}^\dagger(t)\hat{a}(0)\rangle$$
$$= n_c\int_{-\infty}^{+\infty} dt\, e^{i(\Omega-\omega_c)t} e^{-i\langle \delta\hat{\phi}_c^2(t)\rangle/2} = n_c\frac{\gamma_{ST}}{(\Omega-\omega_c)^2 + (\gamma_{ST}/2)^2}. \quad (S17)$$

The photon field has a full-width-half-maximum linewidth $\gamma_{ST}/(2\pi)$. Correspondingly, the coherence time of the maser is $T_{coh} = 2/\gamma_{ST}$. The same coherence time can also be calculated using the magnon phase correlation.

**Note D. Spin-spin correlations**

The spin-spin correlation can be determined from equation (S7), as

$$\langle \hat{S}_+\hat{S}_-\rangle = \frac{w}{2\kappa_S} S_z \left[(1-\frac{\gamma_{eg}}{w})N - (1+\frac{\gamma_{eg}}{w})S_z\right]. \tag{S18}$$

It reaches the maximum value

$$\langle \hat{S}_+\hat{S}_-\rangle = \frac{N^2}{8q}\left(1-\frac{\kappa_c}{2Ng^2 T_2^*}\right)^2, \tag{S19}$$

at pump rate $qw = 2Ng^2/\kappa_c - 1/T_2^*$. The maximum spin correlation is proportional to $N^2$ provided that $2Ng^2 T_2^* \gg \kappa_c$, which is satisfied when the maser operates well above the threshold. Under this condition, the spin ensemble is at half inversion ($S_z = N/2$), and the intra-cavity photon number is $n_c = \kappa_S n_S/\kappa_c = N^2 g^2/(2q\kappa_c^2)$. The incoherent population is mainly the thermal photons ($n_{incoh} \approx n_{th}$) at room temperature ($n_{th} \gg N_e/S_z = 3/2$). Thus, the diffusion coefficient in equation (S16) reduces to $\gamma_{ST} = n_{th}\kappa_c/(2n_c)$, and the minimal diffusion coefficient and maximal coherence time are $\gamma_{ST} = qn_{th}\kappa_c^3/(N^2 g^2)$ and $T_{coh} = 2N^2 g^2/(qn_{th}\kappa_c^3)$, respectively.

**Note E. Sensitivity to external magnetic field and mirror positions**

To estimate the magnetic field sensitivity, we consider a magnetic field noise $\delta B$, which induces a frequency noise of all the NV spins $\delta\omega_S = \gamma_{NV}\delta B$.

The input noise of the cavity can be expressed as $\hat{F}_c(t) = \sqrt{\kappa_c^{ex}}\delta\hat{s}_{in}(t) + \sqrt{\kappa_c^{vac}}\delta\hat{s}_{vac}(t)$, including the cavity internal noise $\delta\hat{s}_{vac}(t)$ and the interaction with the emission channel $\delta\hat{s}_{in}(t)$. The noise operators satisfy the commutation relations $[\delta\hat{s}_{in}(t), \delta\hat{s}_{in}^\dagger(t')] = [\delta\hat{s}_{vac}(t), \delta\hat{s}_{vac}^\dagger(t')] = \delta(t-t')$. Both the internal noise and the emission channel noise contribute to the cavity photon decay, i.e., $\kappa_c = \kappa_c^{ex} + \kappa_c^{vac}$. The output field noise is $\delta \hat{s}_{out}(t) = \delta \hat{s}_{in}(t) - \sqrt{\kappa_c^{ex}}\delta \hat{a}(t)$, the output field is $\hat{s}_{out}(t) = -\sqrt{\kappa_c^{ex}}\hat{a}(t)$, corresponding to output flux $\langle \hat{s}_{out}^\dagger \hat{s}_{out}\rangle = \kappa_c^{ex} n_c$. With small internal cavity losses ($\kappa_c^{vac}/\kappa_c \to 0$), all the cavity decay results in output, and the output power is $P_{out} \approx \hbar \omega_c \cdot \kappa_c n_c$.

In order to calculate the magnetic field sensitivity, we consider the cavity mirrors as fixed with negligible displacement noise. The magnetic field sensitivity can be obtain from the noise spectral density of the output field, $S_{out}(\Omega) = \langle \{-i[\delta \hat{s}_{out}(\Omega) - \delta \hat{s}_{out}^\dagger(\Omega)]\}\{-i[\delta \hat{s}_{out}(-\Omega) - \delta \hat{s}_{out}^\dagger(-\Omega)]\}\rangle$, which gives

$$S_{out}(\Omega) = 1 + \frac{4\kappa_c n_c}{\Omega^2}\left[\frac{\left(\frac{\kappa_S}{2}\right)^2}{\left(\frac{\kappa_c+\kappa_S}{2}\right)^2+\Omega^2} n_{incoh}\frac{\kappa_c}{2n_c} + \frac{\left(\frac{\kappa_c}{2}\right)^2}{\left(\frac{\kappa_c+\kappa_S}{2}\right)^2+\Omega^2}(\gamma_{NV}\delta B \cdot \sqrt{t_m})^2\right], \quad (S20)$$

where the term "1" comes from photon shot noise, the first term in the bracket corresponds to background due to the maser phase fluctuation (linewidth), and the second term in the bracket results from the magnetic field noise. Thus the measured magnetic field sensitivity, is limited by the other two terms, as

$$\delta B \cdot \sqrt{t_m} = \frac{1}{\gamma_{NV}}\frac{\kappa_c+\kappa_S}{\kappa_c}\sqrt{\frac{\Omega^2}{4\kappa_c n_c}\left[1+\frac{4\Omega^2}{(\kappa_c+\kappa_S)^2}\right]+\gamma_{ST}}. \quad (S21)$$

The photon shot noise term is negligible under the slow-noise condition $\Omega \ll \sqrt{2n_{incoh}\kappa_c\kappa_S/(\kappa_c+\kappa_S)}$, $(\kappa_c+\kappa_S)/2$. The magnetic field sensitivity is then determined by quantum diffusion (coherence time) of the maser

$$\delta B \cdot \sqrt{t_m} = \frac{1}{\gamma_{NV}}\frac{\kappa_c+\kappa_S}{\kappa_c}\sqrt{\gamma_{ST}}. \quad (S22)$$

Similarly, the sensitivity to the cavity mirror displacement can be determined by setting the magnetic field as well stabilized. The result of the output noise is

$$S_{\text{out}}(\Omega) = 1 + \frac{4\kappa_c n_c}{\Omega^2} \left[ \frac{(\kappa_S/2)^2}{\left(\frac{\kappa_S + \kappa_c}{2}\right)^2 + \Omega^2} n_{\text{incoh}} \frac{\kappa_c}{2n_c} + \frac{(\kappa_S/2)^2 + \Omega^2}{\left(\frac{\kappa_S + \kappa_c}{2}\right)^2 + \Omega^2} (\frac{\omega_c}{L} \delta x \cdot \sqrt{t_m})^2 \right]. \quad (S23)$$

For low frequency displacement noises [ $\Omega \ll \sqrt{2n_{\text{incoh}}} \kappa_c \kappa_S/(\kappa_c + \kappa_S)$, $\kappa_S/2$ ], the quantum diffusion limited displacement sensitivity is

$$\delta x \cdot \sqrt{t_m} = \frac{L}{\omega_c} \sqrt{n_{\text{incoh}} \frac{\kappa_c}{2n_c}} = \frac{L}{\omega_c} \frac{\kappa_c + \kappa_S}{\kappa_S} \sqrt{\gamma_{\text{ST}}}. \quad (S24)$$

**Note F. Diamond microwave amplifier**

With an input signal $s_{\text{in}} e^{-i\omega_{\text{in}} t}$, the steady-state Langevin equations are

$$0 = wN_g - \gamma_{eg} N_e + ig(a^* S_- - S_+ a), \quad (25a)$$

$$0 = i(\omega_{\text{in}} - \omega_S) S_- - \frac{\kappa_S}{2} S_- + ig S_z a, \quad (25b)$$

$$0 = i(\omega_{\text{in}} - \omega_c) a - \frac{\kappa_c}{2} a - ig S_- + \sqrt{\kappa_{\text{ex}}} s_{\text{in}}, \quad (25c)$$

$$s_{\text{out}} = s_{\text{in}} - \sqrt{\kappa_{\text{ex}}} a. \quad (25d)$$

With the input signal detuning denoted as $\delta_{S,c} = \omega_{\text{in}} - \omega_{S,c}$, the inversion and output are

$$\frac{w - \gamma_{eg}}{w + \gamma_{eg}} = \left\{ 1 + \frac{\frac{2\kappa_{\text{ex}}}{\kappa_c} \frac{4g^2}{\kappa_S \kappa_c} \frac{4|s_{\text{in}}|^2}{w + \gamma_{eg}}}{\left(\frac{4\delta_S^2}{\kappa_S^2} + 1\right)\left(\frac{4\delta_c^2}{\kappa_c^2} + 1\right) - 1 + 2\frac{4\delta_S \delta_c}{\kappa_S \kappa_c} \frac{4g^2}{\kappa_S \kappa_c} S_z + \left(\frac{4g^2}{\kappa_S \kappa_c} S_z - 1\right)^2} \right\} \frac{S_z}{N}, \quad (26a)$$

$$s_{\text{out}} = \frac{\left(1 - \frac{2\kappa_{\text{ex}}}{\kappa_c}\right) - \frac{4\delta_S \delta_c}{\kappa_S \kappa_c} - \frac{4g^2}{\kappa_S \kappa_c} S_z - i \left[\left(1 - \frac{2\kappa_{\text{ex}}}{\kappa_c}\right) \frac{2\delta_S}{\kappa_S} + \frac{2\delta_c}{\kappa_c}\right]}{1 - \frac{4\delta_S \delta_c}{\kappa_S \kappa_c} - \frac{4g^2}{\kappa_S \kappa_c} S_z - i \left(\frac{2\delta_S}{\kappa_S} + \frac{2\delta_c}{\kappa_c}\right)} s_{\text{in}}. \quad (26b)$$

The power gain is $G = |s_{\text{out}}|^2 / |s_{\text{in}}|^2$. For detuning between the cavity, electron spin, and the input signal frequency $\delta_{S,c}/\kappa_{S,c} \sim 1$, the power gain would be significantly reduced to $G \sim O(1)$. However, large power gain is possible under the resonant condition $\delta_{S,c} = 0$.

In the following, we assume the resonant condition. We also assume the cavity loss is mainly caused by the coupling to the input-output channel, i.e. $\kappa_{ex} = \kappa_c$. The inversion reduces to an exactly solvable cubic equation

$$\frac{w-\gamma_{eg}}{w+\gamma_{eg}}N = \left[1 + 2\frac{4g^2}{\kappa_S \kappa_c}\frac{4|s_{in}|^2}{w+\gamma_{eg}}\left(1 - \frac{4g^2}{\kappa_S \kappa_c}S_z\right)^{-2}\right]S_z. \tag{S27}$$

and the power gain reduces to

$$G = \left(1 + \frac{4g^2}{\kappa_S \kappa_c}S_z\right)^2 \Big/ \left(1 - \frac{4g^2}{\kappa_S \kappa_c}S_z\right)^2. \tag{S28}$$

In the amplifying region, we approximately obtain the population inversion for weak input signal $|s_{in}|^2 \ll \frac{g^2(w+\gamma_{eg})}{2\kappa_S \kappa_c}(\frac{w-\gamma_{eg}}{w+\gamma_{eg}}N - \frac{\kappa_S \kappa_c}{4g^2})^2$ as

$$S_z = \frac{w-\gamma_{eg}}{w+\gamma_{eg}}\left[1 - 2\frac{\kappa_S \kappa_c}{4g^2}\frac{4|s_{in}|^2}{w+\gamma_{eg}}\left(\frac{w-\gamma_{eg}}{w+\gamma_{eg}}N - \frac{\kappa_S \kappa_c}{4g^2}\right)^{-2}\right]. \tag{S29}$$

Meanwhile, the power gain is independent of the input signal,

$$G = \left(\frac{w-\gamma_{eg}}{w+\gamma_{eg}}N + \frac{\kappa_S \kappa_c}{4g^2}\right)^2 \Big/ \left(\frac{w-\gamma_{eg}}{w+\gamma_{eg}}N - \frac{\kappa_S \kappa_c}{4g^2}\right)^2. \tag{S30}$$

In the masing region, the approximate inversion for weak input signal $|s_{in}|^2 \ll \frac{w+\gamma_{eg}}{8}(\frac{w-\gamma_{eg}}{w+\gamma_{eg}}N - \frac{\kappa_S \kappa_c}{4g^2})$ is

$$S_z = \frac{\kappa_S \kappa_c}{4g^2}\left(1 - 2\sqrt{\frac{2|s_{in}|^2}{w+\gamma_{eg}}\left(\frac{w-\gamma_{eg}}{w+\gamma_{eg}}N - \frac{\kappa_S \kappa_c}{4g^2}\right)^{-1}}\right), \tag{S31}$$

and the power gain depends on the input signal (nonlinear amplification) is

$$G = \left(\sqrt{\frac{w+\gamma_{eg}}{2|s_{in}|^2}\left(\frac{w-\gamma_{eg}}{w+\gamma_{eg}}N - \frac{\kappa_S \kappa_c}{4g^2}\right)} - 1\right)^2. \tag{S32}$$

The output power is independent of the weak input signal in the masing region (red star in Fig.3d in the main text).

Figure S1 shows the inversion, the power gain, and the noise temperature as functions of the pump rate for a weak input signal ($P_{in} = 1$ fW) and a fixed microwave cavity $Q$ facor ($Q = 10^5$). For pump below the threshold ($w < \gamma_{eg}$), the system is in the absorbing region ($G_{dB} < 0$ dB). For pump above the threshold ($w > \gamma_{eg}$), the system works as a microwave amplifier until reaching the masing region, which ceases when over-pumping occurs at $w > w_{max} = \left(4g^2 N/\kappa_c - 2/T_2^*\right)/q$. Note that in the masing region, both stable and unstable solutions exist. The power gain significantly increases near the maser threshold and the noise temperature is at sub-Kelvin level.

Figure S2 shows dependence of the power gain on the cavity $Q$ factor, the pump rate, and the input signal power.

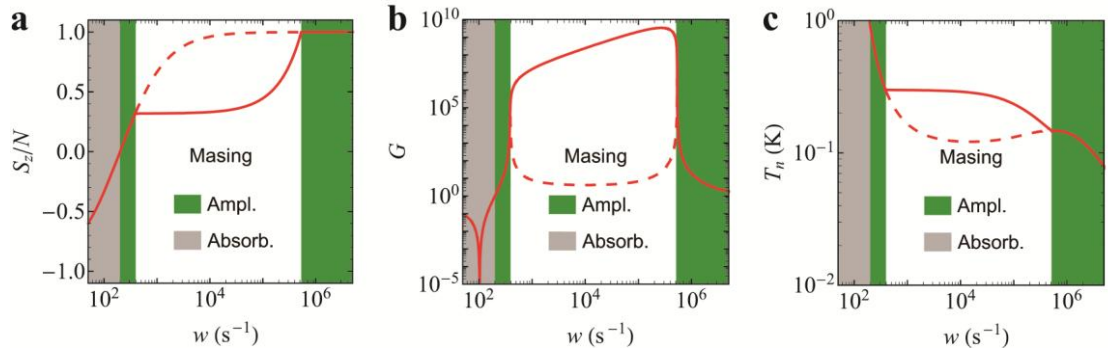

**Figure S1 | Pump rate dependence of room-temperature diamond microwave amplifier.** The inversion (**a**), power gain (**b**), noise temperature (**c**) as functions of the pump rate $w$ for a weak input signal $P_{in}$=1 fW and a fixed $Q$=10$^5$. The absorbing, amplifying and masing regions are marked as grey, green and white. The solid lines are stable solutions and the dashed lines are the unstable solutions in the masing region.

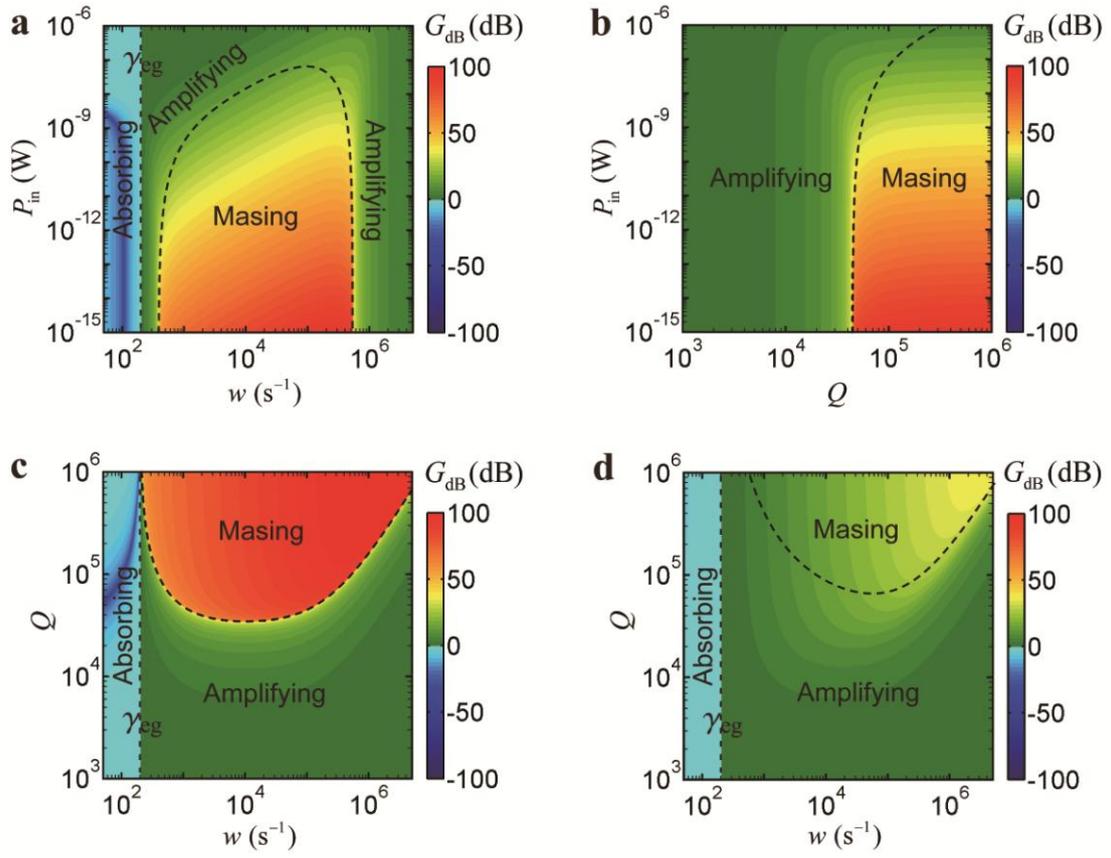

**Figure S2| Dependence of power gain of room-temperature diamond microwave amplifier on the pump rate, cavity loss, and input power.** The power gain in decibel for (**a**) fixed $Q = 10^5$, (**b**) fixed $w = 10^5$ s$^{-1}$, (**c**) fixed input signal power $P_{in} = 1$ fW, and (**d**) fixed input power $P_{in} = 10$ nW. The dashed black curve implies the maser threshold. The other parameters are the same as Fig. 2 in the main text, i.e., $\omega_{in}/2\pi = \omega_c/2\pi = \omega_S/2\pi = 3$ GHz, $g/2\pi = 0.02$ Hz, $T_2^* = 0.5$ μs, $N = 0.375 \times 10^{14}$, $T = 300$ K, and $\gamma_{eg} = 200$ s$^{-1}$.